\documentclass[preprint,12pt]{elsarticle}

\newcommand{\beq}{\begin{eqnarray}}
\newcommand{\eeq}{\end{eqnarray}}
\newcommand{\bit}{\begin{itemize}}
\newcommand{\eit}{\end{itemize}}
\newcommand{\eps}{\varepsilon}

\newcommand{\om}{\omega}
\newcommand{\p}{\partial}
\newcommand{\de}{\delta}

\newcommand{\ekT}{\frac{e}{k_BT}}
\newcommand{\np}{n^+}
\newcommand{\nm}{n^-}
\newcommand{\dm}{\lambda_m}

\usepackage{graphicx}
\usepackage{pstricks}
\usepackage{amssymb}
\usepackage{amsmath}
\usepackage{amsfonts}

\journal{Advances in Planar Lipid Bilayers and Liposomes}

\newcommand{\kp}{\mathbf{k}_\perp}
\newcommand{\kpm}{k_\perp}
\newcommand{\rp}{\mathbf{r}_\perp}

\newcommand{\epsm}{\epsilon_m}

\begin{document}

\begin{frontmatter}




\author[ESPCI,UF,ICS]{F. Ziebert}
\author[ESPCI]{D. Lacoste}

\ead{david.lacoste@gmail.com}
\cortext[Cor]{Tel: +33140795140; Fax: +33140794731}
\address[ESPCI]{Laboratoire de Physico-Chimie Th\'eorique - UMR CNRS Gulliver 7083,
ESPCI, 10 rue Vauquelin, F-75231 Paris, France}
\address[UF]{Physikalisches Institut, Albert-Ludwigs-Universit\"at, 79104 Freiburg, Germany}
\address[ICS]{Institut Charles Sadron, 23 rue du Loess, 
67034 Strasbourg Cedex 2, France}

\title{A planar lipid bilayer in an electric field: membrane instability, flow field and electrical impedance}


\begin{abstract}
For many biotechnological applications it would be useful
to better understand the effects produced by electric fields on lipid membranes.
This review discusses several aspects
of the electrostatic properties of a planar lipid membrane
with its surrounding electrolyte in a normal DC or AC electric field.

In the planar geometry, the analysis of electrokinetic equations
can be carried out quite far, allowing to characterize
analytically the steady state and the dynamics of the charge accumulation in the Debye layers,
which results from the application of the electric field.
For a conductive membrane in an applied DC electric field, we
characterize the corrections to the elastic moduli,
the appearance of a membrane undulation instability
and the associated flows which are built up near the membrane.
For a membrane in an applied AC electric field, we analytically derive the impedance
from the underlying electrokinetic equations.
We discuss different relevant effects due to the membrane conductivity or due to
the bulk diffusion coefficients of the ions.
Of particular interest is the
case where the membrane has selective conductivity for only one type of ion.
These results, and future extensions thereof, should be useful
for the interpretation of impedance spectroscopy data
used to characterize e.g. ion channels embedded in planar bilayers.
\end{abstract}


\end{frontmatter}


\section{Introduction}

Bilayer membranes formed from phospholipid molecules are
an essential component of the membranes of cells. The mechanical properties
of equilibrium membranes are characterized by two elastic moduli,
the surface tension and the curvature modulus \cite{seifert_mb_review:1997}, which typically
depend on the electrostatic properties of the membranes and its surroundings \cite{andelman}.
Understanding how these properties are modified when the membrane is driven
out of equilibrium is a problem of considerable importance to the
physics of living cells.
A membrane can be driven out of equilibrium in many ways, for instance
by ion concentration gradients or by electric fields.

Quite generally one can distinguish between
systems in which the electric field is applied externally and
systems which are able to self-generate electric fields:

\subsection{Membranes in externally applied electric fields}
The external application of electric fields on lipid films is used to
produce artificial vesicles
(electroformation), as well as to create holes in the
membrane (electroporation) \cite{electroporation:1989}.
Both processes are important for biotechnological
applications and they are widely used experimentally.
However, they are still not well understood theoretically.
The research on electroformation is motivated by the
hope to produce artificial lipid vesicles in a controlled
and simple way, which will be key to many biotechnological
applications.
Cell electroporation is a popular technology and
biomedical applications of in vivo cell electropermeabilization \cite{vernier:2006}
are gaining momentum for drug and nucleic acids electrotransfer
and for the destruction of tumor cells for cancer treatment \cite{Mir:2000}.

In view of the importance of these applications,
many research efforts have been
devoted to study and understand deformations of giant unilamellar vesicles (GUVs)
due to the application of electric fields.
In the presence of an AC electric field, giant unilamellar vesicles
show a rich panel of possible behaviors and morphological transitions
depending on experimental conditions -- electric field frequency, conductivities of the
medium and of the membrane, salt concentration, etc. \cite{lipowsky:08,dimova:09}.
A theoretical framework involving hydrodynamics and a continuum mechanics
description of the membrane has been developed, which accounts quantitatively for
the observed equilibrium and non-equilibrium shapes taken by the vesicles
in the presence of an AC electric field \cite{petia:2009,svetina:2007}.
For a clear and self-contained presentation of this theoretical framework,
we recommend  the chapter "Non-equilibrium dynamics of lipid membranes:
deformation and stability in electric fields" by P. Vlahovska, in the same issue of this book.

The application of external fields is also interesting as a means
to move fluids via electro-osmosis \cite{Armand:2000,Ramos:2000}
and to self-assemble colloidal particles, for various technological applications.
Moreover, the ability to move fluids and nanoparticles
at small scales is used in many biological systems.
For instance, membrane-bound ion pumps and channels are able to transport
water (for instance in aquaporin channels) and ions (in ionic pumps and channels)
in a particularly selective and efficient way, which one would like
to reproduce in artificial or biomimetic microfluidic devices.

\subsection{Membranes in self-generated electric fields}
In some cases of biological relevance, membranes are able to self-generate an electric field,
due to embedded ion channels or pumps. This can be achieved
because the channels are able to transport ions from one side of the membrane
to the other in a selective way, either down their
concentration gradient in passive transport or against it in active transport, e.g.~
coupled to the hydrolysis of Adenosine triphosphate or activated by light.
Probably the best known example is the opening and closing of ion channels in nerve cells
allowing the transmission of an electric signal via action potentials \cite{hille}.
For all these reasons, ion channels and pumps play an essential
role in many biological functions of a cell \cite{cell}.

In order to better understand how nerve cells operate in vivo, it would be helpful to construct an in vitro
biomimetic equivalent which would have some key features of the in vivo system, such as the ability to
generate an action potential, but without the complexity of a real nerve cell.
Active membranes, which
are giant unilamellar vesicles containing
ion pumps such as bacteriorhodopsin \cite{jb_PRL,jb_PRE,Faris2009}
are a promising system to achieve this goal.

\vspace{3mm}
The main purpose of this review is to propose and analyze
a simple model 
to foster the understanding of various
effects resulting from electric fields acting on a planar lipid membrane.
Although we are mostly interested in applications to biological or biomimetic systems
composed of lipid membranes, we would like to point out that the theoretical framework
presented here is very general. It can be easily adapted to analyze 
the electrical properties of artificial membranes which can have
very different properties from biological membranes
(as far as e.g.~ionic conductivities or the bending stiffness are concerned).

This review is organized as follows: in section \ref{sec:mb_instability} we
present the model
for a planar lipid membrane and its surrounding fluid in an applied electric field.
In this section, we will restrict ourselves to
the case of a DC field. In particular we will focus on
i) the electrostatic and electrokinetic steady-state corrections
to the elastic moduli of the membrane due to the application of the electric field,
see section \ref{growthrate};
ii) the flow fields which can be predicted from such an
approach, at steady state and in the case that the membrane is ion-conductive, see
section \ref{sec:flow_fields}.
In section \ref{comp_exp} we will compare the model predictions to two relevant experiments.
More details on this theoretical framework, as well as an extension
to the nonlinear electrostatic regime
using the Poisson-Boltzmann equation,
can be found in Refs.~\cite{Lacoste2007,Lacoste2009,ziebert:2010,ziebert_NJP}.
Finally, in section \ref{sec:impedance} we present an analysis of
the model in the presence of
time-dependent AC electric fields. We provide
derivations for the impedance of the system from the underlying electrokinetic equations,
for situations where the membrane is either blocking or selectively conductive for ions.

\section{A quasi-planar membrane in a DC electric field}
\label{sec:mb_instability}

The mechanical properties of membranes at equilibrium are characterized by
two elastic moduli, the surface tension and the bending modulus.
These moduli typically depend on electrostatic properties,
and their modifications in the case of charged membranes or surfaces in an electrolyte
have been examined theoretically in various situations:
in the linearized Debye-H\"uckel approximation as well as
in the nonlinear Poisson-Boltzmann (PB) regime,
for lipid monolayers and symmetric bilayers
\cite{andelman,Helf88,Lekk89,WinterHelf92}.
More recently, charged asymmetric bilayers with unequal Debye lengths
on both sides of the membrane \cite{chou:1997}
and an uncharged membrane in a DC field \cite{lomholt_elect} have been investigated.

In all the works mentioned above, a free energy approach has been used.
Note that while this method works well for equilibrium membranes,
it is not applicable to situations in which the membrane fluctuations
have a non-equilibrium origin, as in the case of active membranes
containing ion channels \cite{PB,RTP,jb_PRL,jb_PRE}
or in the case of a membrane in a time dependent electric field.
In our recent work \cite{Lacoste2007,Lacoste2009,ziebert:2010,ziebert_NJP},
we thus have studied
this problem using an electrokinetic approach, which does not have
the limitations of a free energy formulation.
In this framework we allow for a finite conductivity of the membrane
due to e.g.~ion channels or pumps, and the ion transport is
described using a Poisson-Nernst-Planck approach \cite{hunter,kumaran:2000,leonetti:2004}.
The electrostatic corrections to the elastic moduli and the fluid flows in the electrolyte
are then obtained by imposing
the overall force balance at the membrane.

Two additional points are worth emphasizing: first,
our approach is able to correctly describe the capacitive effects
of the membrane and of the Debye layers
while keeping the simplicity of the ''zero-thickness approximation''
on which most of the literature on lipid membranes is based.
This is accomplished by the use of an effective Robin-type
boundary condition (BC) at the membrane.
Second, as the method is based on a calculation of the general force balance at the membrane,
additional non-equilibrium processes could be included into the model rather easily.
For simplicity we investigate here only
the effects of ionic currents flowing through the membrane,
which in turn affect the fluid flow near the membrane. Other
non-equilibrium effects that could be included as well are for instance
including ion channel stochasticity or active pumping.

\subsection{Model equations - electrostatics}
\label{eq_estat}

\begin{figure}[t!]
\begin{minipage}{0.5\textwidth}
\includegraphics[width=.85\textwidth]{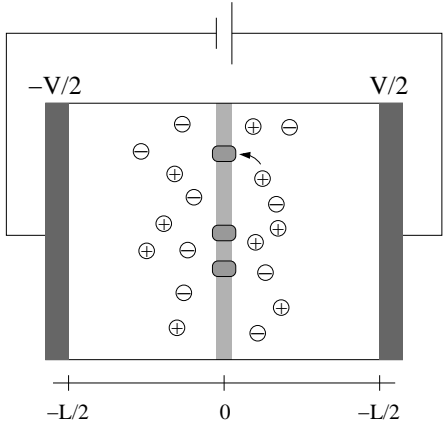}
\end{minipage}
\begin{minipage}{0.5\linewidth}
\caption{\label{fig_sketch}
Sketch of a quasi-planar membrane embedded in a symmetric electrolyte. The
initially flat bilayer membrane is represented by the plane $z=0$.
The membrane fluctuations around this base state have not been represented.
A voltage $V$ is applied far from the membrane on electrodes separated by a distance $L$.
The membrane carries ion channels which give rise to a conductance G.
}
\end{minipage}
\end{figure}

Figure \ref{fig_sketch} shows a sketch of the planar geometry that is studied:
we consider a steady current driven by a DC voltage drop $V$ across
two electrodes separated by a fixed distance $L$.
The membrane is quasi-planar and located at $z=0$.
It is embedded in an electrolyte of monovalent ions with number densities $n^+$
and $n^-$. It contains channels for both ion species but is itself neutral,
i.e. does not carry fixed charges.
The channels or pumps are assumed to be homogeneously distributed in the membrane
and enter only in the effective conductance $G$, as introduced below.
A point in the membrane is characterized in the Monge representation
by the height function $h(\rp)$, where $\rp$ is a two-dimensional in-plane vector.
The base state of this problem is a flat membrane.
Hence the electric field, assumed to be perfectly aligned in $z$-direction,
is perpendicular to it.
We assume a quasi-static approach \cite{lomholt_elect,Lacoste2009}
in which membrane fluctuations are much slower than the characteristic diffusion time
$\tau=\frac{1}{D\kappa^2}$ for the ions to diffuse a Debye length.

In the electrolyte, the electric potential $\phi$ obeys Poisson's equation
\beq
\label{Poisson_n}
\nabla^2\phi=-\frac{1}{\epsilon}\left(en^+ - en^-\right)=-\frac{2}{\epsilon}\rho\,.
\eeq
Here $e$ is the elementary charge,
$\epsilon$ is the dielectric constant of the electrolyte
and we have introduced {\it half} of the charge density,
\beq\label{rhodef}
\rho=e\frac{\np-\nm}{2}\,.
\eeq
For the sake of simplicity, we assumed a symmetric $1:1$ electrolyte, thus
far away from the membrane $n^+=n^-=n^*$, and
the total system is electrically neutral. The densities of the ion species
obey the Poisson-Nernst-Planck (PNP) equations
\beq\label{PNP}
\p_t n^\pm+\nabla\cdot\mathbf{j}^{\pm}=0\,\,,\,\,\,\,\mathbf{j}^{\pm}=D\left(-\nabla n^{\pm}\mp n^{\pm}\frac{e}{k_B T}\nabla\phi\right)\,,
\eeq
where $\mathbf{j}^\pm$ are the particle current densities of the ions
and $k_B T$ is the thermal energy.
We will assume here that both ion types
have the same diffusion coefficient $D$. Note that we will discuss the effects
of differing diffusion coefficients for an applied AC voltage
in section \ref{D_uneq}.

Since we are primarily interested in the behavior close to the membrane,
for the boundary conditions (BC) far away from the membrane we assume
\beq
\label{VatL}
\phi\left(z=\pm L/2\right)&=&\pm V/2\,,\\
\label{rhoatL}
\rho\left(z=\pm L/2\right)&=&0\,.
\eeq
Eq.~(\ref{VatL}) states that
the potential at the electrodes
is held fixed externally. This BC is quite oversimplified for real electrodes,
but captures the main effects of the electric field,
see the discussion in Ref.~\cite{ziebert:2010}.
We have also assumed that the distance between the electrodes is much larger
than the Debye length, $L\gg\lambda_D=\kappa^{-1}$, where
\beq
\kappa=\sqrt{\frac{2e^2n^*}{\epsilon k_B T}}=\lambda_D^{-1}\,.
\eeq
Hence, as already mentioned above, the bulk electrolyte is quasi-neutral
with negligible charge density (compared to the total salt concentration)
and far from the membrane Eq.~(\ref{rhoatL}) holds.

The BC at the membrane is crucial to correctly account for capacitive effects.
We use the Robin-type BC (see \ref{app:RobinBC} for a derivation)
\beq\label{RobinBC}
&&\dm(\mathbf{n}\cdot\nabla)\phi_{|z=h^+}=\dm(\mathbf{n}\cdot\nabla)\phi_{|z=h^-}=\phi(h^+)-\phi(h^-)\,,
\eeq
where $\mathbf{n}$ is the unit vector normal to the membrane and
\beq\label{dmdef}
\dm=\frac{\epsilon}{\epsilon_m}d\,.
\eeq
$\lambda_m$ is a length scale
containing the membrane thickness $d$ and the ratio
of the dielectric constants $\epsilon/\epsilon_m$  of the electrolyte and the membrane.
Note that in Eq.~(\ref{RobinBC}), the membrane plays a similar role as the Stern layer in the description of
Debye layers near a charged interface. This BC was rederived for electrodes
sustaining Faradaic current \cite{itskovich1977,bazant2005}
or charging capacitively \cite{bazant2004PRE}, and was applied for membranes
in Refs.~\cite{leonetti:2004,Lacoste2009,ziebert:2010}. There it was
shown to properly account for the jump in the charge distribution which occurs near
the membrane as a result of the dielectric mismatch between the membrane and the surrounding electrolyte.

In addition to Eq.~(\ref{RobinBC}), we impose the
continuity of the bulk current $j^\rho_{|z=0}$ at the membrane.
This BC involves the ohmic law
\beq
j^\rho_{|z=0}=-\frac{G}{e}[\mu^\rho]_{z=0}\,,
\eeq
where $G$ denotes the membrane conductance per area and $\mu^\rho$ the
electrochemical potential.
The electrostatic potential and the ion densities can now be obtained by solving
Eq.~(\ref{Poisson_n}) in the linear
Debye-H\"uckel approximation  and one obtains \cite{ziebert:2010}:

\noindent i) the jump of the charge density at the membrane, $\rho_m$,

\noindent ii) the current through the membrane, $j_m$, and

\noindent iii) the electric field inside the membrane, $E_0^m$:
\beq\label{a_formula}
\rho_m&=&\frac{\frac{\epsilon\kappa^2}{2}V-\frac{j_m}{D}(L+\dm)}{2+\kappa\dm}\,,\\
\label{sigma_form}
j_m&=&-j^\rho=\frac{GV}{1+\frac{2}{\epsilon\kappa^2 D}GL}\,,\\
\label{intern_field}
E_0^m&=&-\frac{1}{d}\left[\frac{2}{\eps\kappa^2}\left(-\frac{j_m L}{D}-2\rho_m\right)+V\right]\,.
\eeq
For simplicity, in the derivation of Eqs.~(\ref{a_formula}-\ref{intern_field})
we assumed equal ion conductivities ($G^+=G^-=G$) and a symmetric electrolyte on both sides
of the membrane ($\kappa^{>0}=\kappa^{<0}=\kappa$). Note that
the method presented in this section can
be easily extended to cover more general cases.
In additon, the nonlinear electrostatic problem
(keeping the Poisson-Boltzmann equation) can be still solved
analytically in the non-conductive case. The nonlinear generalizations of
Eqs.~(\ref{a_formula}-\ref{intern_field}) can be found in Ref.~\cite{ziebert_NJP}.

\subsection{Model equations - hydrodynamics and force balance at the membrane}

The hydrodynamics of the electrolyte is described by the
incompressible Stokes equation, $-\nabla p+\eta\nabla^2 \mathbf{v}+\mathbf{f}=0$ with $\nabla\cdot \mathbf{v}=0$,
where $\mathbf{v}$ is the velocity field of the electrolyte,
$\eta$ its viscosity, $p$ the hydrostatic pressure and
 $\mathbf{f}=-2\rho\nabla\phi$ the electric driving force.
From the solution of the electrostatic and the hydrodynamic problem, one obtains
 the total stress tensor
\beq\label{stresstensor}
\tau_{ij}=-p\delta_{ij}+\eta\left(\p_i v_j+\p_j v_i\right)
+\epsilon\left(\hspace{-1mm}E_i E_j -\frac{1}{2}\delta_{ij}E^2\hspace{-1mm}\right),\,\,\,
\eeq
which contains the pressure, the viscous stresses in the fluid
and the Maxwell stresses.

The lipid bilayer membrane, on the other hand,  behaves as a two dimensional fluid
which can store elastic energy in bending deformations. More precisely,
its elastic properties can be described by the standard
Helfrich free energy
\beq \label{mb free energy1}
F_H=\frac{1}{2} \int d^2 \rp [ \Sigma_0 \left( \nabla h \right)^2 + K_0 \left( \nabla^2 h
\right)^2  ],
\eeq
where $\Sigma_0$ is the bare surface
tension
and $K_0$ the bare bending modulus of the membrane.

All forces present in the system, the electrostatic, viscous and elastic ones,
have to fulfill the force balance equation. The latter states
that the discontinuity of the
normal-normal component of the stress tensor, as defined
in Eq.~(\ref{stresstensor}) and evaluated at the membrane position,
must equal the restoring force due to membrane's elasticity, hence
\beq
\label{BC_stress_normal}
-\left(\tau_{zz,1|z=h^+}-\tau_{zz,1|z=h^-}\right)
= -\frac{\partial F_H}{\partial h(\rp)}
=\left(-\Sigma_0\kpm^2-K_0\kpm^4\right)h(\kp)\,.\,\,\,
\eeq
Here the index $1$ in the stress tensor refers to the order of an
expansion with respect to the membrane height field $h(\rp)$.
Note that at zeroth order, the membrane is flat and
thus only electric forces and osmotic pressure balance.
By expanding to linear order in the height field $h(\rp)$, and using
\beq
h\propto h_0 e^{i\kp\cdot\rp +s(\kp)t}\,,
\eeq
Eq.~(\ref{BC_stress_normal}) yields
the growth rate $s(\kp)$ of membrane fluctuations.
Details of the derivation of $s(\kp)$ can be found
in Refs.~\cite{Lacoste2007,Lacoste2009,ziebert:2010}.
We would like to emphasize that the force localized
at the membrane surface is a priori unknown in this problem. Thus it
must be determined self-consistently from the BCs for the velocity and the stress.

\subsection{Growth rate and renormalized elastic moduli}
\label{growthrate}

The force balance Eq.~(\ref{BC_stress_normal})
determines the growth rate $s(\kpm)$ entering the normal stress difference,
\beq\label{disp_fin}
\eta \kpm s(\kpm)&=&
-\frac{1}{4}\left(\Sigma_0+\Delta\Sigma\right)\kpm^2-\Gamma_\kappa\kpm^3
-\frac{1}{4}\left(K_0+\Delta K\right)\kpm^4\,.\quad
\eeq
The electrostatic corrections to the surface tension,
$\Delta\Sigma=\Delta\Sigma_\kappa+\Delta\Sigma_m$, and to
the bending modulus, $\Delta K=\Delta K_\kappa+\Delta K_m$
can be decomposed into:

\noindent i) an outside contribution due to the charges
accumulated in the Debye layers and denoted with the index $\kappa$;

\noindent ii) an inside contribution due to the voltage
drop at the membrane and denoted with an index $m$.
They are given by
\beq\label{deltasigma}
\Delta\Sigma_\kappa&=&-4\frac{\rho_m^2}{\epsilon\kappa^3}-16\frac{\rho_m j_m}{\epsilon\kappa^4 D}\,\,,\,\,\,\,
\Delta K_\kappa=\frac{3\rho_m^2}{\epsilon\kappa^5}
\eeq
for the contribution due to the Debye layers and by
\beq\label{deltasm}
\Delta\Sigma_m&=&-\epsm (E^m_0)^2 d\,\,,\,\,\,\,
\Delta K_m=\epsm (E^m_0)^2 \left(\frac{d^3}{12}-\frac{\rho_m}{E^m_0}\frac{d}{\eps\kappa^3}\right)
\eeq
for the contribution due to the field inside the membrane.

\begin{figure}[t!]
\begin{minipage}{0.45\textwidth}
\includegraphics[width=0.95\textwidth]{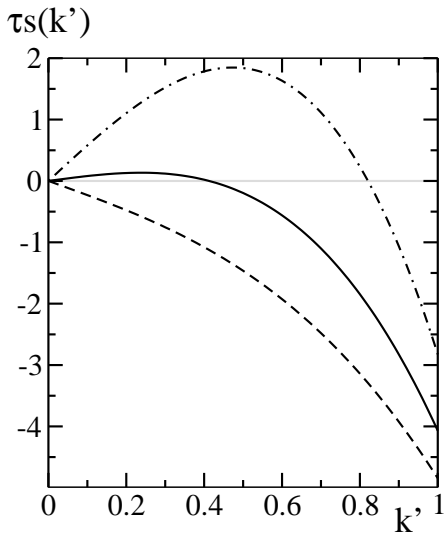}
\end{minipage}
\hspace{2mm}
\begin{minipage}{0.50\linewidth}
\caption{\label{dispfig}
    The renormalized growth rate or dispersion relation, $\tau s$,
    as a function of the rescaled wave number $k'=\kpm/\kappa$ for three voltages:
    $V=0.7{\rm V}$ (dashed line),
    $V=0.75{\rm V}$ (solid line), $V=0.8{\rm V}$ (dash-dotted line).
    We have used the following parameters: dielectric constants
    $\epsilon=80\epsilon_0$ and $\epsilon_m=2\epsilon_0$;
    membrane thickness $d=5{\rm nm}$ leading
    to $\dm=\frac{\epsilon}{\epsilon_m}d =200{\rm nm}$;
    diffusion coefficient of ions $D=10^{-9}{\rm m}^{2}{\rm s}^{-1}$;
    viscosity $\eta=10^{-3}{\rm Pa}\,{\rm s}$; inverse Debye length
    $\kappa=2\cdot 10^{7}{\rm m}^{-1}$; bare surface tension $\Sigma_0=1$mNm$^{-1}$;
    bare bending modulus $K_0=10k_B T$. Here we assumed a non-conductive membrane,
    $G=0$.
    }
\end{minipage}
\end{figure}

Note that in Eq.~(\ref{disp_fin}), one also obtains a purely non-equilibrium correction
$\Gamma_\kappa=\frac{4\rho_m j_m}{\epsilon\kappa^5 D}$.
It would correspond to a term proportional to $k_\perp^3$ in an ''effective
membrane free energy'' incorporating the Maxwell stresses.
At equilibrium such a term is forbidden by symmetry,
but in a non-equilibrium situation, where the membrane sustains a current $j_m\neq0$,
it is allowed. For realistic parameters, however, this term is very small, see
Ref.~\cite{Lacoste2009} for a detailed discussion.

The inside contribution to the membrane surface tension is always negative, see Eq.~(\ref{deltasm}). The same is typically
true for the outside contribution, see Eq.~(\ref{deltasigma}) and note that $\rho_m,j_m>0$.
Hence these contributions can overcome the bare surface tension $\Sigma_0$.
If this is the case, an
instability towards membrane undulations sets in.
Such an instability had already been described
for the high salt limit in Ref.~\cite{pierre}.
Note that the linearized theory developed here
describes only the early stages of the instability, but it is more general
than previous works since it is not limited to the high salt limit and
in addition accounts for hydrodynamic effects.
The linear growth rate of the membrane fluctuations given by
Eq.~(\ref{disp_fin}) is shown in Fig.~\ref{dispfig} in rescaled units.
We scaled the wave vector by
$\kappa$, hence $k'=\kpm/\kappa$ and the time by the typical time for
ions to diffuse a Debye length, $\tau=\frac{1}{D\kappa^2}$.
The control parameter of the instability is the external voltage $V$.
Fig.~\ref{dispfig} shows the growth rate for three different levels of the voltage:
the dashed line is for $V=0.7{\rm V}$, which lies below the threshold of the instability,
all wave numbers are damped and the membrane is stable.
The solid and the dash-dotted line correspond to $V=0.75{\rm V}$ and $0.8{\rm V}$.
These values are above threshold and the growth rate is positive for
a finite wave number window.

For a more detailed discussion of the dependance of the corrections
to the elastic moduli, the instability threshold and the characteristic wave number
as a function of salt concentration and membrane conductivity,
we refer the reader to Refs.~\cite{Lacoste2009,ziebert:2010}.

\subsection{Flow fields near a driven membrane}
\label{sec:flow_fields}

\begin{figure}[t!]
\begin{minipage}{0.5\textwidth}
\includegraphics[width=.9\textwidth]{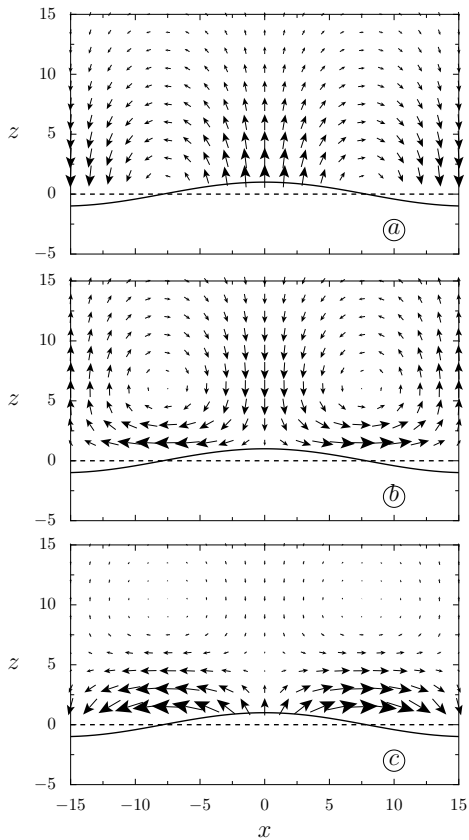}
\end{minipage}
\begin{minipage}{0.5\linewidth}
\caption{\label{flowfig} Representation of the flows around the membrane
    beyond the instability threshold. The orientation
    of the electric field is towards negative values of $z$.
    Panel a) shows the flow generated by the membrane bending mode.
    Panel b) shows the ICEO flow.
    Finally, panel c) shows the actual flow,
    which is the superposition of the former two
    and results in a strong flow near the membrane, oriented
    parallel to the surface. Both axes are scaled by the Debye length $\kappa^{-1}$.
    Parameters are as in Fig.~\ref{dispfig} except $V=3.165{\rm V}$,
    $\kappa=10^{7}{\rm m}^{-1}$, $G=10 S{\rm m}^{-2}$ and $L=10\mu{\rm m}$.
    }
\end{minipage}
\end{figure}

We now summarize the main features of the fluid flows which arise near the
membrane when it is driven by ionic currents \cite{Lacoste2009}.
Fig.~\ref{flowfig} was generated by selecting the fastest
growing wave number and using the corresponding maximum growth rate.
The shape of the membrane undulation is represented with the black solid curves.
Fig.~\ref{flowfig}c) shows the flow field for a high membrane conductance and low
salt, in the regime where the membrane is unstable due to the
electrostatic correction to the surface tension and thus starts to
undulate.  The resulting flow is a superposition of two distinct flows:
first, the typical flow associated to a membrane bending mode
\cite{BrochardLennon} as shown in Fig.~\ref{flowfig}a). Second, the flow
which results from the ion transport across the membrane. The latter flow has
 the typical counter-rotating vortices of an ICEO flow \cite{bazant:2004}, as shown in Fig.~\ref{flowfig}b).
Clearly, the superposition of these two flow contributions, Fig.~\ref{flowfig}c),
results in a parallel flow close to the membrane, in contrast to the usual bending
mode flow given by Fig.~\ref{flowfig}a).

For most realistic parameters -- modest conductivities, not too low salt --
the flow generated by membrane bending is usually dominating
and hides the small ICEO contribution. To be able to observe
the flows of Fig.~\ref{flowfig}, a high membrane conductance $G$ and low salt are needed.
Also, since for macroscopic electrode distances
$L$ (e.g.~of the order of millimeters),
the voltage needed to induce the instability is very high,
we have assumed a microscopic electrode distance of $L=10\mu{\rm m}$.
While it might still be possible to observe flows for higher
salt and macroscopic electrode separations, such situations can
not be analyzed within the Debye-H\"uckel approximation used here.

Note that somewhat similar looking flow patterns have been recently observed
experimentally in vesicles subject to AC electric fields in Ref.~\cite{lipowsky:08}.
On closer inspection, however, it appears that these flows most probably
have a different origin from the ICEO flows, since they are more likely to result
from electrophoresis of charged lipids within the membrane.

\subsection{Applications to specific experiments}
\label{comp_exp}

Here we will briefly discuss how the framework presented above can be applied
to recent experiments:
the first experiment studied supported membranes subject to an electric field \cite{charitat_EPJE},
while the second one investigated active membranes \cite{jb_PRL,jb_PRE,Faris2009}.

S. Lecuyer {\it et al.} \cite{charitat_EPJE}
recently performed neutron reflectivity measurements on a system consisting of two nearby
membrane bilayers in an external AC electric field. One of the bilayers was
close to the bottom electrode and used to protect the second one from
interacting with the wall. The bare values of the elastic moduli
were known from X-ray off-specular experiments
for a similar system \cite{charitat_PNAS}, yielding
$\Sigma_0 \simeq 0.5 $mNm$^{-1}$ and $K_0 \simeq 15 k_BT$.
The experiments were performed in an
AC electric field at several frequencies.
For the lowest frequency ($10{\rm Hz}$) and for a voltage of $V=5$V,
the electrostatic corrections to the surface tension and bending modulus
were found to be $\Delta \Sigma \simeq -1 \pm 0.15$mNm$^{-1}$ and $\Delta K \simeq 185 \pm 15 k_BT$.

Assuming that the membrane is non-conductive, $G=0$,
and using an inverse Debye length of $\kappa=2\cdot10^7{\rm m}^{-1}$ (milli-Q water)
and the experimental electrode distance of $L=1{\rm mm}$,
our model yields
$\Delta\Sigma\simeq-2\cdot $mNm$^{-1}$ and $\Delta K\simeq190 k_B T$.
Thus the model successfully accounts for the
order of magnitude of the electrostatic corrections observed in this experiment.
Note, however, that the linearized Debye-H\"uckel approach is not a good approximation in this case,
as applied voltages are rather high. For this reason, we
recently extended our work to the Poisson-Boltzmann regime \cite{ziebert_NJP}.


The second experiments we would like to discuss concerns active membranes,
which are artificial lipid vesicles containing bacteriorhodopsin ionic pumps
\cite{jb_PRL,jb_PRE,Faris2009}. These pumps are able to transfer protons
unidirectionally across the membrane
by undergoing light-activated conformational changes.
The transport of protons across the membrane eventually builds up
a transmembrane potential.
In Refs.~\cite{PB,RTP,jb_PRE}, a hydrodynamic theory has been
developed to describe the nonequilibrium fluctuations of the
membrane induced by the activity of the pumps.
This work triggered substantial theoretical interest in the problem,
mainly focusing on the proper description
for these non-equilibrium effects associated with protein conformational
changes \cite{gov:2004,gautam,Lacoste2005,chen,lomholt_mb_actives}.

In these models for active membranes, the electrostatic effects
associated with the ion transport were not explicitly described.
The framework presented in this review provides a more detailed
description of the ion transport, which could be useful to understand
some aspects of active membrane experiments.
From a contour analysis of giant active vesicles, the fluctuation
spectrum of the membrane was measured in Ref.~\cite{Faris2009},
and a lowering of the membrane tension produced by the activity of the pumps was reported.
Only the correction to the surface tension
has been accurately measured in this experiment and many aspects of
the transport of ions are still unknown.
However, for simplicity we can assume that the passive state corresponds
to a non-conductive membrane, $G=0$, and the active state to a membrane with
conductance $G=10S$m$^{-2}$.
If we also assume a typical transmembrane potential of the order of $50 {\rm mV}$,
we can use the results for the corrections to the surface tension obtained above.
Accounting for the rather high amount of salt using $\kappa \simeq 5 \cdot 10^8$m$^{-1}$,
we find a reasonable
estimate for the observed tension lowering, $\Delta\Sigma\simeq3 \cdot 10 ^{-7}$Nm$^{-1}$.
We also find that there is no measurable difference for the bending modulus between the active
and passive state, in agreement with the experiments. The model further predicts
a current density of $j_m \simeq 1$Am$^{-2}$ when
the pumps are active, which corresponds to an overall current of $1$pA on a vesicle of size 1$\mu$m.

This accord in orders of magnitude for the electrostatic corrections
is quite promising.
For a more detailed comparison between experiments and the presented model,
it would be necessary to do experiments in varying conditions
(ionic strength, conductance of the membrane, or orientation of the pumps in the membrane for instance).
Combined measurements of the
membrane current and the transmembrane potential in the same experiment,
using e.g.~patch-clamp techniques, would also be desirable.

\section{Impedance of a planar membrane in an AC electric field}
\label{sec:impedance}

Impedance spectroscopy \cite{bockris} is an effective tool to obtain a 
characterization of the electric properties of lipid bilayer membranes.
The method has been used in particular for supported lipid bilayers,
which are a promising experimental system to characterize membrane proteins, channels
or inclusions
and more generally constitute the basis of highly sensitive
detection technologies, i.e. biosensors \cite{sackmann:2002}.
In the recent work \cite{steinem:2009}, for instance, impedance spectroscopy
has been used to characterize
gramicidin D channels in pore suspending membranes.
Nowadays, many biotechnology companies develop systems to measure the impedance
of whole cells for e.g.~screening or drug delivery.

In many cases, the interpretation of the data obtained by impedance spectroscopy
is not that straightforward.
Typically one uses equivalent circuits, which are sometimes controversial,
since different models can be used for fitting the data.
Moreover, there is often a lack of knowledge concerning
the conditions of validity of these equivalent circuits to describe
the diffuse charging in electric Debye layers.
To answer these questions, one possibility is to start with an electrokinetic description
based on the Poisson-Nernst-Planck equations. With such an approach,
the dynamics of diffuse charging \cite{bazant2004PRE} and
the current voltage relation in electrochemical thin films have been successfully
analyzed \cite{bazant2005}.
This approach is also useful for relating impedance measurements
to the properties of the diffuse layers
near charge selective interfaces such as electrodes or
ion-exchange membranes \cite{zalzman:2009}.

In the following, we extend the model
studied in the previous sections to the case of an applied AC electric field.
For simplicity the membrane will be assumed to be {\it strictly planar and non-fluctuating}.
We use the Poisson-Nernst-Planck equations to evaluate the impedance of this system, which
can be then compared to simple equivalent circuits.
We will first present the generic time-dependent equations for the perturbation induced
by the applied AC field. Then we proceed to calculate the impedance for the following cases:
i) an ideally blocking membrane with equal diffusion coefficients for the two ion species,
ii) the same system but with unequal ion diffusion coefficients
and finally iii) an ideally non-blocking membrane which conducts selectively only one type of ion.

\subsection{Time-dependent electric fields}
The Poisson-Nernst-Planck equations for an electrolyte
have already been given in section \ref{eq_estat}.
Taking the time derivative of the Poisson equation, Eq.~(\ref{Poisson_n}), one
obtains
\beq
-\epsilon \p_t \nabla^2 \phi= e (\p_t n^+ - \p_t n^-)= -e ( \nabla\cdot\mathbf{j}^+ - \nabla\cdot\mathbf{j}^-),
\eeq
where in the last equation, the conservation of ion densities, Eqs.~(\ref{PNP}), has been used.
Through integration over space (assuming a one dimensional geometry), and using the
definition of the electric field, $\mathbf{E}=-\nabla \phi$, it
follows that \cite{jackson,gaspard_ion_transport}
\beq
\label{total current}
\mathbf{I}=\epsilon \p_t \mathbf{E} + e\mathbf{J},
\eeq
where the constant in of integration, $\mathbf{I}$, is the total electric current density.
The first term on the r.h.s.~in Eq.~(\ref{total current})
is the displacement current. The second term,
$\mathbf{J}=\mathbf{j}^+ - \mathbf{j}^-=2 \mathbf{j_\rho}$, is the particle current density.
The displacement current was absent in the previous section
because we assumed a stationary state, but for the time-dependent case
it is crucial to obtain 
the response to an externally
applied AC electric potential.
We note that by virtue of the Poisson equation, Eq.~(\ref{Poisson_n}),
the total current density is divergence-free,
$\nabla \cdot \mathbf{I} = 0$, at all times. Further note
that the experimentally measurable quantity is
given by the total electric current. For this reason, it is the relevant
quantity to calculate impedance as shown below.

\subsection{Equations for time-periodic perturbations of an equilibrium base state}

Let us assume an established
equilibrium solution $c_0^+(z)$, $c_0^-(z)$ and $\phi_0(z)$ for the electrolyte
in the absence of the AC field, which could be caused
by an additional DC field or a Nernst potential.
For convenience we consider here the {\it charge} densities $c^{\pm}$. Note that $c^{\pm}=en^{\pm}$ and
$\kappa^2=2ec_0/(\epsilon k_B T)$.
The equations for the electrostatic problem, see Eqs.~(\ref{Poisson_n}, \ref{PNP})
above, read
\beq
\epsilon\p_z^2\phi&=&c^- - c^+\,,\\
\p_tc^{\pm}&=&-\p_zj^{\pm}\,,\\
j^\pm&=&-D^\pm\left(\p_zc^\pm \mp c^\pm\frac{e}{k_BT}\p_z\phi\right)\,.
\eeq
Linearization around the base state like
\beq
c^+=c_0^+ + \eta C^+\,,\,\,
c^-=c_0^- + \eta C^-\,,\,\,
\phi=\phi_0+\eta \Phi\,,\nonumber
\eeq
where $\eta$ is a small book-keeping parameter,
leads at order $\mathcal{O}(\eta^0)$ to
\beq
c_0^\pm=c_0e^{\mp\frac{e\phi_0(z)}{k_BT}},\,\,
\,{\rm with}\,\,\phi_0\,\,{\rm solution\,\,of}\,\,\,\,\,
\epsilon\p_z^2\phi_0&=c_0\left(e^{\phi_0}-e^{-\phi_0}\right)\,.\nonumber
\eeq
This restates that the equilibrium solution has to fulfill the classical PB equation.
At order $\mathcal{O}(\eta^1)$ in the perturbations, we get
\beq
\label{PBal}\epsilon\p_z^2\Phi&=&C^- - C^+\,,\\
\label{PNPp}\p_tC^\pm&=&D^\pm\p_z\left(\p_zC^\pm \mp c_0^\pm{\ekT}\p_z\Phi \mp C^\pm{\ekT}\p_z\phi_0\right)\,.
\eeq
As already discussed in the general case above, taking the time-derivative
of Eq.~(\ref{PBal}), insertion of the linearized PNP equations (\ref{PNPp})
and integration in $z$ yields
\beq
\epsilon\p_z\p_t\Phi&-&D^-\left(\p_zC^- -c_0^-\ekT\p_z\Phi -C^-\ekT\p_z\phi_0\right)\nonumber\\
&+&D^+\left(\p_zC^+ +c_0^+\ekT\p_z\Phi +C^+\ekT\p_z\phi_0\right)=I(t)\,.\nonumber
\eeq
The integration constant $I(t)$ is the total electric current density.
As we are interested in the response to an AC external voltage, $V(t)=V_0e^{i\om t}$,
introducing $I(t)=I_0e^{i\om t}$ and $\Phi\propto e^{i\om t}$, we arrive at
\beq\label{Pbal_int}
&&\left(i\om\epsilon+\ekT(D^+c_0^+ + D^-c_0^-)\right)\p_z\Phi\nonumber\\
&&\quad\quad\quad+D^+\p_zC^+-D^-\p_zC^-+\left(D^+C^+ +D^-C^-\right)\ekT\p_z\phi_0=I_0\,.\quad\,\,\,
\eeq
The first term on the l.h.s.~is the displacement current.
The remaining terms are currents due to concentration gradients
and a current induced by the equilibrium potential at the membrane.
All these contributions taken together yield the total current $I_0$
in response to the external AC field.

We are left with the problem to
solve Eqs.~(\ref{PNPp}) and (\ref{Pbal_int}) with the external voltage entering via the boundary conditions,
just like in section \ref{sec:mb_instability}.


\subsection{Impedance for an ideally blocking non-conductive membrane}

The equations derived in the last section are
general as they describe the first order perturbation in an electrolyte
induced by an AC voltage externally imposed at some boundaries.
Let us now apply them to the planar membrane geometry
as sketched in Fig.~\ref{fig_sketch}. The membrane is assumed to be flat and located
at $z=0$. The AC voltage will be externally applied at the electrodes at $z=\pm L/2$.
For simplicity, we assume that there is no additional DC electric field or Nernst potential,
i.e. that the equilibrium solution is given by the homogeneous solution
$\phi_0=0,\, c_0^{\pm}=c_0$.

First we will treat the simplest case of an ideally blocking, non-conductive
membrane, $j^\pm(0)=0$. We also assume equal diffusion coefficients for the
positive and negative ions, $D^+=D^-=D$.
Then the above equations (\ref{PNPp}, \ref{Pbal_int}) for the perturbations reduce to
\beq
\label{ib1}
&&i\om C^\pm=D\p_z\left(\p_zC^\pm \mp c_0{\ekT}\p_z\Phi\right),\\
\label{ib3}&&\left(i\om\epsilon+2Dc_0\ekT\right)\p_z\Phi+D(\p_zC^+-\p_zC^-)=I_0\,.
\eeq
Due to the symmetry of our system, one has
\beq\label{symm_geom}
\Phi(z,t)=-\Phi(-z,t)\,,\,\,\,\rho(z,t)=-\rho(-z,t)\,,\,\,\,c(z,t)=c(-z,t)\,.
\eeq
Hence it is enough to solve the problem in $z\in[-L/2,0]$.
The BCs in the chosen geometry read
\beq
\label{cp0}&&C^+(-L/2)=0\,,\\
\label{cm0}&&C^{-}(-L/2)=0\,,\\
\label{potV}&&\Phi(-L/2)=-V_0/2\,,\\
\label{jp0}&&\p_zC^+(0)+c_0\ekT\p_z\Phi(0)=0=j^+(0)/D\,,\\
\label{jm0}&&\p_zC^-(0)-c_0\ekT\p_z\Phi(0)=0=j^-(0)/D\,,\\
\label{robin}&&\lambda_m\p_z\Phi(0)=\Phi(0^+)-\Phi(0^-)\,.
\eeq
Eqs.~(\ref{cp0}-\ref{potV}) fix the densities and the potential at the electrodes,
as has already been discussed in section \ref{eq_estat}.
The next two equations (\ref{jp0}, \ref{jm0}) state that the membrane is non-conductive
for both ion types.
Finally the last equation (\ref{robin}) is again the Robin-type BC
describing the capacitive behavior of the membrane
with the effective length scale $\lambda_m=\frac{\epsilon}{\epsilon_m}d$.
We will use the first five BCs to fix the five integration constants of
Eqs.~(\ref{ib1}, \ref{ib3}).
Then imposing the last condition will yield the
current-voltage relation and finally the impedance.

Extracting an equation for $C_s=C^++C^-$ by adding Eqs.~(\ref{ib1}) 
yields $i\om C_s=D\p_z^2 C_s$. From the BCs $\p_zC_s(0)=0=C_s(-1/2)$ it follows $C_s(z)=0$,
i.e. the total density of particles (positively and negatively charged)
remains homogeneous.
Introducing $\rho=C^+-C^-$
and substracting Eqs.~(\ref{ib1}) 
yields
\beq\label{iomrho}
&&i\om\rho=D\p_z^2\rho+D\epsilon\kappa^2\p_z^2\Phi\,,\\
\label{pot_rho}
&&\left(i\om\epsilon+D\epsilon\kappa^2 \right)\p_z\Phi+D\p_z\rho=I_0\,,
\eeq
where we have used $\frac{c_0 e}{k_B T}=\epsilon\kappa^2/2$.
Eq.~(\ref{pot_rho}) can be integrated, yielding
\beq
\Phi(z)=c_1+\frac{I_0 z-D\rho(z)}{D\epsilon\kappa^2+i\om\epsilon}\,.\nonumber
\eeq
The BCs $\rho(-L/2)=0$, $\Phi(-L/2)=-V_0/2$ fix the constant of integration to
\beq
c_1=-\frac{V_0}{2}+\frac{I_0 L/2}{D\epsilon\kappa^2+i\om\epsilon}\,.\nonumber
\eeq
Insertion of the obtained potential into Eq.~(\ref{iomrho}) for $\rho$ yields
\beq\label{rhoeq2}
\frac{D\kappa^2+i\om}{D}\rho=\p_z^2\rho\,.
\eeq
Using again the obtained potential transforms the BC
$\p_z\rho(0)+\epsilon\kappa^2\p_z\Phi(0)=0$
into the simpler form $\p_z\rho(0)=i\frac{I_0\kappa^2}{\om}$. Together
with $\rho(-L/2)=0$,
the solution of Eq.~(\ref{rhoeq2}) can be given as
\beq\label{rhosol}
\rho(z)=i\frac{I_0\kappa^2}{\beta\om\cosh(\beta L/2)}\sinh\left[\beta(z+L/2)\right]
\eeq
with the (complex) inverse length scale
\beq
\beta=\sqrt{\kappa^2+i\om/D}\,.
\eeq
The remaining BC, Eq.~(\ref{robin}),
is a jump condition at the membrane.
What we have calculated above are the solutions $\Phi^{<0}$, $\rho^{<0}$ on $z\in[-L/2,0]$.
Using the symmetry of our problem, Eq.~(\ref{symm_geom}), one
directly obtains $\Phi^{>0}$, $\rho^{>0}$ on $z\in[0,L/2]$.
Imposing Eq.~(\ref{robin}), $\lambda_m\p_z\Phi(0)=\Phi^{>0}(0)-\Phi^{<0}(0)$, then yields
\beq
\lambda_m \frac{I_0-D\p_z\rho(0)}{D\epsilon\kappa^2+i\om\epsilon}
&=&c_1^{>0}+\frac{-D\rho^{>0}(0)}{D\epsilon\kappa^2+i\om\epsilon}
-\left[c_1^{<0}+\frac{-D\rho^{<0}(0)}{D\epsilon\kappa^2+i\om\epsilon}\right]\nonumber\\
&=&-2c_1^{<0}-2\frac{-D\rho^{<0}(0)}{D\epsilon\kappa^2+i\om\epsilon}\,. \nonumber
\eeq
Solving for the external voltage $V_0$ -- note that it enters in the integration constant
$c_1$ of the electric potential -- one gets
\beq
V_0&=&\frac{I_0 L}{D\epsilon\kappa^2+i\om\epsilon}-2\frac{D\rho(0)}{D\epsilon\kappa^2+i\om\epsilon}
+\lambda_m \frac{I_0-D\p_z\rho(0)}{D\epsilon\kappa^2+i\om\epsilon}\,.\nonumber 
\eeq
This is the current-voltage relation. The impedance is defined as
$Z(\om)=V(\om)/AI(\om)=V_0/(A I_0)$, with $A$ the membrane area normal to the $z$-direction.
Using the expression for the density, Eq.~(\ref{rhosol}), one arrives at the following expression
for the impedance of a non-conductive membrane
\beq\label{Z_brute}
Z=\frac{L/A}{D\epsilon\kappa^2+i\om\epsilon}
-i\frac{(L/A)\frac{D\kappa^2}{\om}}{D\epsilon\kappa^2+i\om\epsilon}
\frac{\tanh\left[\beta(L/2)\right]}{\beta L/2}
+\frac{(\lambda_m/A) \left(1-i\frac{D\kappa^2}{\om}\right)}{D\epsilon\kappa^2+i\om\epsilon}\,.\,
\eeq

Let us discuss the obtained result. The first term is the contribution of the electrolyte.
This can be seen by rewriting it as
\beq
Z_B=\frac{1}{R_B^{-1}+i\om C_B}
\eeq
and identifying the capacitance of the bulk, $C_B=\epsilon A/L$, which is in parallel with
the resistance of the bulk $R_B=\frac{1}{D\epsilon\kappa^2}\frac{L}{A}=\frac{L k_B T}{2D c_0 e A}$.
A similar interpretation holds for the term
$\frac{(\lambda_m/A)}{D\epsilon\kappa^2+i\om\epsilon}$
in Eq.~(\ref{Z_brute}), which can be written as
\beq
Z_S=\frac{1 }{ R_S^{-1}+i\om C_S}\,.
\eeq
This is again a RC-circuit
with the capacitance $C_S=\epsilon A/\lambda_m=\epsilon_m A/d$ of the membrane and a resistance
$R_S=\frac{1}{D\epsilon\kappa^2}\frac{\lambda_m}{A}$.
It arises from the Robin-BC which involves the effective length scale $\lambda_m$
defined in Eq.~(\ref{dmdef}).
One can thus recast Eq.~(\ref{Z_brute}) into the form
\beq\label{Zfull_nc1d}
Z=Z_B+Z_S
-\frac{i}{\om} \frac{D\kappa^2}{R_B^{-1}+i\om C_B}
\left[\frac{\tanh\left[\beta(L/2)\right]}{\beta L/2}
+\frac{\lambda_m}{L}\right]\,.
\eeq
The last term in this equation, let us call it $Z_C$,
is due to charging of the double layer and the membrane.
This can be best seen in the limit $\lambda_D/L=1/(\kappa L)\ll1$, i.e.
when the Debye length is small compared to the system size.
Then $\frac{\tanh\left[\beta(L/2)\right]}{\beta L/2}\simeq2/(\kappa L)$
and in the prefactor, the resistance $R_B^{-1}$ dominates over the capacitance.
One gets
\beq\label{ZCdef}
Z_C
\simeq\frac{1}{i\om C_{eff}}\,,
\eeq
with the effective capacitance $C_{eff}=\epsilon\frac{A}{2\lambda_D+\lambda_m}$.
Note that the thickness of the corresponding planar capacitor is the sum of the
two Debye layers thicknesses ($2\lambda_D$)
and the effective length $\lambda_m$ describing the capacitive effects of the membrane.

\begin{figure}[t]
\centering
\includegraphics[width=0.6\textwidth]{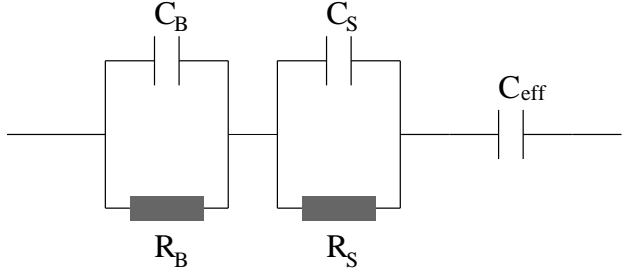}
\caption{\label{fig_ers1}
Effective circuit for the ideally blocking non-conductive membrane, Eq.~(\ref{Zfin_nonc_1D}):
Two RC-circuits, one for the bulk and one for the membrane are in series
with the effective charging capacitance of the membrane.
}
\end{figure}

As shown in Fig.~\ref{fig_ers1}, for the blocking non-conductive membrane
one effectively has an association in series of
the RC-circuit of the bulk, the RC-circuit
of the membrane and the effective capacitance of the charging membrane
\beq\label{Zfin_nonc_1D}
Z=Z_B+Z_S+Z_C=\frac{1}{R_B^{-1}+i\om C_B}+\frac{1 }{ R_S^{-1}+i\om C_S}+\frac{1}{i\om C_{eff}}\,,
\eeq
as long as $\lambda_D/L\ll1$ holds.
As $\lambda_m\simeq 200 {\rm nm}$, the impedance contribution $Z_S$
is usually small compared to the bulk resistance
and can be neglected for $L\gg\lambda_m$.
However, the contribution described by $\lambda_m$ to the charging impedance $Z_C$
can be of similar order as the one from the Debye layers
and might even dominate the charging.

\begin{figure}[t]
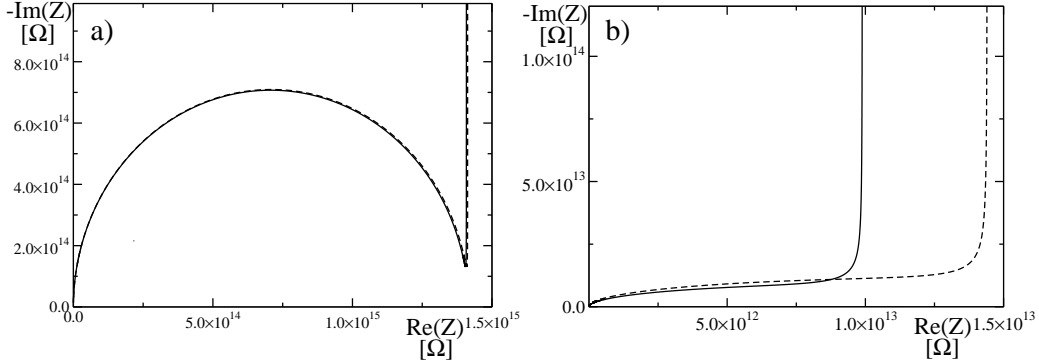

\centering
\includegraphics[width=0.5\textwidth]{fig5a.eps}
\includegraphics[width=0.5\textwidth]{fig5b.eps}
\caption{\label{fig_block1}
Panel a) shows a Nyquist plot for a macroscopic system size, $L=1$mm.
At low frequency the behavior is governed by the charging of the membrane.
The semi-circle is governed by the bulk-RC-circuit.
As $\lambda_D/L\simeq10^{-3}$, the effective circuit and the full
calculation agree well.
Panel b) shows a Nyquist plot for a microscopic system, $L=10\mu{\rm m}$.
As $L$ decreases, the bulk becomes less important and the
RC semi-circle less pronounced.
The full calculation (solid curve) yields a lower resistance
for the charging process at low frequencies
than the effective circuit (dashed curve).
Parameters as in Fig.~\ref{dispfig} except for $\kappa=10^{-6}{\rm m}^{-1}$ (pure water);
membrane area $A=1\mu{\rm m}^2$.
}
\end{figure}

The best way to visually present the impedance is by a so-called Nyquist plot \cite{bockris}.
There one traces the negative imaginary part, $-{\rm Im}[Z(\om)]$, of the impedance
as a function of its real part, ${\rm Re}[Z(\om)]$, for varying frequency $\om$.
Nyquist plots for the full impedance, Eq.~(\ref{Zfull_nc1d}), and for the limit
$\lambda_D/L\ll1$, Eq.~(\ref{Zfin_nonc_1D}), are shown in Fig.~\ref{fig_block1}.
Panel a) shows the case of a macroscopic system size, $L=1$mm.
One clearly notices the RC-semi-circle terminating for high frequencies at the
origin. For the given parameters one enters this semi-circle at $\omega\simeq50{\rm Hz}$;
the maximum is achieved for $\om_{RC}=D\kappa^2=1{\rm kHz}$.
The low frequency branch is dominated by the membrane charging
capacitively at $R\simeq R_B+R_S$, thus for low frequencies
one has a divergence like $(i\om C_{eff})^{-1}$.
As $\lambda_D/L\simeq10^{-3}$, the effective circuit and the full
calculation agree well.
Fig.~\ref{fig_block1}b) shows the case of a microscopic system size, $L=10\mu{\rm m}$.
Here the bulk RC-signal is much less pronounced and charging dominates entirely.
The full calculation (solid curve) yields a lower resistance
for the charging process at low frequencies
than the effective circuit obtained by the
small-Debye layer approximation (dashed curve).

\subsection{Non-conductive membrane: effect of unequal diffusion coefficients}
\label{D_uneq}

In this section we investigate the effect of differing diffusion coefficients
for the two ion species,
$D^+\neq D^-$, on the impedance of a blocking non-conductive membrane.
Except for this assumption, the calculation is analogous to the one of the previous section.
Equations (\ref{PNPp}), (\ref{Pbal_int}) for the perturbations now read
\beq
\label{ib1D}
&&i\om C^\pm=D^\pm\p_z\left(\p_zC^\pm \mp \frac{\epsilon\kappa^2}{2}  \p_z\Phi\right),\\
\label{ib3D}&&\left(i\om\epsilon+(D^++D^-) \frac{\epsilon\kappa^2}{2} \right)\p_z\Phi+D^+\p_zC^+-D^-\p_zC^-=I_0\,.
\eeq
The BCs are still given by Eqs.~(\ref{cp0}-\ref{robin}).
Since the equations for the charge densities do not decouple as before,
it is useful to introduce $C=C^++C^-$ and $\rho=C^+-C^-$ again, yielding
\beq
&&i\om C=\p_z^2\left[\bar D C + \de \rho +2\de\bar c\Phi \right],\nonumber\\
&&i\om \rho=\p_z^2\left[\de C + \bar D \rho +2\bar D\bar c\Phi \right],\nonumber\\
&&\left(i\om\epsilon+2\bar D\bar c\right)\p_z\Phi+\de\p_zC+\bar D\p_z\rho=I_0\,.\nonumber
\eeq
Here we introduced the average and the difference of the two diffusion coefficients
\beq
\bar D=(D^++D^-)/2\,,\,\,\,\delta=(D^+-D^-)/2\,.
\eeq
Integration of the equation for the potential $\Phi$ yields
\beq
\Phi(z)=c_1+\frac{I_0 z-(\de C(z)+\bar D\rho(z))}{\bar D\epsilon\kappa^2+i\om\epsilon}
\,\,\,\,{\rm with}\,\,\,\,
c_1=-\frac{V_0}{2}+\frac{I_0 L/2}{\bar D\epsilon\kappa^2+i\om\epsilon}\,.\nonumber
\eeq
Insertion into the equations for $C$ and $\rho$ yields a matrix equation
\beq\label{matrixeq}
\left( \begin{array}{cc}
i\om - \left[\bar D-\de^2\epsilon\kappa^2/N\right]\p_z^2 & -\left[\de i\om\epsilon/N\right]\p_z^2 \\
-\left[\de i\om\epsilon/N\right]\p_z^2 & i\om - \left[\bar D i \om\epsilon/N\right]\p_z^2
\end{array} \right)
\cdot
\left(
\begin{array}{c}
\hspace{-1mm}C\hspace{-1mm}  \\ \hspace{-1mm}\rho\hspace{-1mm}
\end{array}
\right)
=0\,,
\eeq
where we have introduced $N=\bar D\epsilon\kappa^2+i\om\epsilon$.

Assuming solutions of the form
$C,\rho\propto e^{\beta z}$,
Eq.~(\ref{matrixeq}) yields four solutions for the decay length $\beta$.
In the case of equal diffusion coefficients studied previously, $\delta=0$ and the equations are decoupled.
In that case $\bar D=D$ and one simply gets $\beta_1^2=\frac{i\om}{D}$ associated to the relaxation
of the total particle density $C$
and $\beta_2^2=\frac{N}{D\epsilon}=\frac{D\kappa^2+i\om}{D}$ associated to the
relaxation of $\rho$, see Eq.~(\ref{rhoeq2}).
In the case of unequal diffusion coefficients,
the equations are coupled and the general solutions
are 
\beq\label{fullbetaeq}
\beta_{1,2}^2&=&\frac{i\om\bar D+\left({\bar D}^2-\delta^2\right)\kappa^2/2
\mp\sqrt{(\kappa^2/2)^2\left({\bar D}^2-\delta^2\right)^2-\de^2\om^2}}
{\left({\bar D}^2-\delta^2\right)}\,.\,\, 
\eeq
Here the minus sign applies to $\beta_1$ and the plus sign to $\beta_2$.
Consequently, Eq.~(\ref{matrixeq}) is solved by the ansatz
\beq\label{gen_sol_2D}
\left(
\begin{array}{c}
\hspace{-1mm}C\hspace{-1mm}  \\ \hspace{-1mm}\rho\hspace{-1mm}
\end{array}
\right)
=
\sum_{i=1,2}\bigg[A_i
\left( \begin{array}{c}
\hspace{-1mm}E_i\hspace{-1mm}  \\
\hspace{-1mm}1 \hspace{-1mm}
\end{array} \right)
\sinh\left[\beta_i\left(z+L/2\right)\right] 
+B_i
\left( \begin{array}{c}
\hspace{-1mm}E_i\hspace{-1mm}  \\
\hspace{-1mm}1\hspace{-1mm}
\end{array} \right)
\cosh\left[\beta_i\left(z+L/2\right)\right]
\bigg]\nonumber
\eeq
with the eigenvectors given by
\beq
E_i=\frac{\left[\de i\om\epsilon/N\right]\beta_i^2}
{i\om - \left[\bar D-\de^2\epsilon\kappa^2/N\right]\beta_i^2}\,.\nonumber
\eeq
The effective BCs read:  
$\p_zC(0)=0$ and $\p_z\rho(0)=  i\frac{2\bar c I_0}{\om\epsilon}$ at $z=0$;
$C(-L/2)=0$ and $\rho(-L/2)$ at $z=-L/2$.
The last two BCs yield $E_1B_1+E_2B_2=0$ and $B_1+B_2=0$.
As $E_1\neq E_2$ this implies $B_1=0=B_2$, i.e. the $\cosh$-contributions in the solution vanish.
After some algebra one obtains
\beq
\rho&=&i\frac{I_0\kappa^2}{\om} \frac{E_1 E_2}{E_2-E_1}
\left[
\frac{\sinh\left[\beta_1\left(z+L/2\right)\right]}{E_1\beta_1\cosh(\beta_1L/2)}
-\frac{\sinh\left[\beta_2\left(z+L/2\right)\right]}{E_2\beta_2\cosh(\beta_2L/2)}
\right],\nonumber\\
C&=&i\frac{I_0\kappa^2}{\om} \frac{E_1 E_2}{E_2-E_1}
\left[
\frac{\sinh\left[\beta_1\left(z+L/2\right)\right]}{\beta_1\cosh(\beta_1L/2)}
-\frac{\sinh\left[\beta_2\left(z+L/2\right)\right]}{\beta_2\cosh(\beta_2L/2)}
\right].\nonumber
\eeq
Using the Robin-type BC, Eq.~(\ref{robin}), 
and once again the symmetry of the problem
one gets
\beq
\lambda_m \frac{I_0-\left(\de\p_zC(0)+\bar D\p_z\rho(0)\right)}{2\bar D\bar c+i\om\epsilon}
=-2c_1^{<0}-2\frac{-\left(\de C^{<0}(0)+\bar D\rho^{<0}(0)\right)}{2\bar D\bar c+i\om\epsilon}\,.\nonumber     
\eeq
Solving for $V_0$, insertion of the obtained solutions for $C$ and $\rho$ and applying
$Z=V_0/(I_0 A)$ one obtains the impedance
\beq\label{Z2Dbrute}
Z&=&
\frac{L/A}{\bar D\epsilon\kappa^2+i\om\epsilon}
+\frac{(\lambda_m /A)\left(1-\bar D i\frac{\kappa^2}{\om}\right)  }
{\bar D\epsilon\kappa^2+i\om\epsilon} \nonumber\\
&&-\,i\,\frac{2\kappa^2/A}{(\bar D\kappa^2+i\om)\om\epsilon}
\frac{E_1 E_2}{E_2-E_1}\left(\delta+\frac{\bar D}{E_1}\right)
\frac{\tanh\left[\beta_1L/2\right]}{\beta_1}\nonumber\\
&&+\,i\,\frac{2\kappa^2/A}{(\bar D\kappa^2+i\om)\om\epsilon}
\frac{E_1 E_2}{E_2-E_1}\left(\delta+\frac{\bar D}{E_2}\right)
\frac{\tanh\left[\beta_2L/2\right]}{\beta_2}.
\eeq
The first two contributions are already familiar to us, they stem
from the bulk and the Stern-like description of the membrane.
Note that $\bar D$ enters instead of $D$.

Let us discuss the newly arising terms.
As an expansion in $\lambda_d/L\ll 1$ is a bit tedious, let us consider only
the simpler limit $\kappa\rightarrow\infty$.
Eq.~(\ref{fullbetaeq}) for $\beta_1^2$ has a minus sign in front of the square root,
the two $\kappa$-terms cancel and
\beq\label{beta1Warb}
\beta_1^2=\frac{i\om\bar D}{\left({\bar D}^2-\delta^2\right)}\,\,\rightarrow\,\,\beta_1=\pm\sqrt{i\om/D_{eff}}\,
\eeq
with $D_{eff}=\left({\bar D}^2-\delta^2\right)/\bar D$.
For $\beta_2^2$ one has the plus sign in front of the square root, the terms in $\kappa^2$ dominate
and one simply gets $\beta_2=\pm\kappa$.
For the eigenvectors to leading order one has
$E_1=\frac{\left({\bar D}^2-\de^2\right)\kappa^2}{\de i \om}\,\,,\,\,E_2=-\frac{1}{E_1}$
and
$\frac{E_1 E_2}{E_2-E_1}\left(\delta+\frac{\bar D}{E_1}\right)
=\frac{i\om\delta^2}{\kappa^2\left({\bar D}^2-\de^2\right)}$,
$\frac{E_1 E_2}{E_2-E_1}\left(\delta+\frac{\bar D}{E_2}\right)=-\bar D$.

Consequently, the last term in Eq.~(\ref{Z2Dbrute}) exactly reduces to
the Debye-layer part of the
charging contribution. 
Finally one obtains at leading order in $\lambda_D$
\beq\label{ZWfin}
Z=\bar{Z}_B+\bar{Z}_S+Z_C+Z_W
\eeq
with
\beq\label{warburg_2D}
Z_W=\frac{2\lambda_D^2}{\bar D\epsilon A/L}
\frac{\delta^2}{\left({\bar D}^2-\de^2\right)}
\frac{\tanh\left[\beta_1L/2\right]}{\beta_1 L}.
\eeq
The first two terms are the RC-contributions of the bulk and the membrane
(note that $\bar D=(D^++D^-)/2$ enters instead of $D$).
The charging capacitance $Z_C$ of the membrane is unchanged.
The last term is the so-called Warburg impedance,
with $\beta_1=\sqrt{i\om/D_{eff}}$ and
$D_{eff}=\left({\bar D}^2-\delta^2\right)/\bar D$.
Note that this contribution is only present for unequal diffusion
coefficients $\de=(D^+-D^-)/2\neq0$.
It is proportional to $\lambda_D^2$ at leading order
\footnote{For simplicity, we used the limit $\kappa\rightarrow\infty$
to derive this term. Taking this limit strictly, the contribution would vanish --
as then both charge types diffuse infinitely rapidly across the zero-thickness Debye-layer.
In real systems, however, $\kappa$ remains always finite and thus one should include
the leading order contribution, $Z_W$, in the impedance.}.

\begin{figure}[t]
\centering
\includegraphics[width=0.7\textwidth]{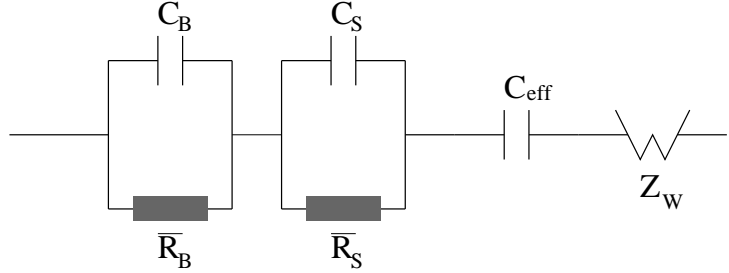}
\caption{\label{fig_ers2}
Effective circuit for the ideally blocking non-conductive membrane with differing
diffusion coefficients,
Eq.~(\ref{ZWfin}):
Two RC-circuits, one for the bulk and one for the membrane are in series
with the effective charging capacitance
and a Warburg resistance.
}
\end{figure}

The effective circuit
corresponding to the obtained impedance is shown in Fig.~\ref{fig_ers2}.
The contribution $Z_W$ has been first described by Warburg \cite{warburg1,warburg2}
for electrochemical systems; in a nutshell, it arises from damped
concentration oscillations
close to an interface, here the membrane.
We note however, that with typical differences in diffusion coefficients $D^+/D^-=0.1..10$,
a Nyquist plot of Eq.~(\ref{ZWfin}) is indistinguishable from
Fig.~\ref{fig_block1} obtained for
equal diffusion coefficients.
This is due to the fact that in the geometry under investigation,
the charging of the membrane is highly dominating
the low-frequency behavior as it is proportional to $\om^{-1}$.
Nevertheless,  experiments often display a Warburg-like impedance at
low frequencies, see e.g.~Ref.~\cite{steinem:2009}.
In the next section we will investigate the case of a slightly
conductive ion-selective membrane and will find that in this case
one indeed obtains a Warburg impedance. We thus postpone a discussion
of $Z_W$ to the next section.

\subsection{Impedance for an ideally non-blocking conductive membrane}
\label{Zconduct}

For many applications it is interesting to account for a
small but non-zero membrane conductivity. This is important for instance
in the context of  the
characterization of ion channel proteins or pumps embedded in a
lipid membrane using impedance spectroscopy.
In contrast to section \ref{eq_estat}, where we discussed the effects of a DC voltage
on a conductive membrane that lets pass both types of charged ions ($G^+=G^-=G$),
here we will treat the case of a {\it selective membrane}, which lets
pass only the positive ions. Thus, we assume a
linearized relation $j^+=G^+\Delta\mu^+$ where $G^+$ is the effective conductance
per unit area. The negative ions are not allowed to pass the membrane, hence $j^-=0$
or effectively $G^-=0$.
This situation is relevant for biomembranes,
where ion channels allow the passage of positively charged ions
like ${\rm Na}^+$ or ${\rm K}^+$, but not of negatively charged ions
like ${\rm Cl}^-$ which are typically larger.

To simplify the analysis, we will not describe the structure of
the Debye layers as explicitly as in the previous sections.
Instead we rely on two
known approximations used in the study of electrochemical systems:

i) the bulk is to a good approximation {\it locally} electroneutral.
More precisely, deviations from electroneutrality
occur only in the third order in an expansion of $\lambda_D/L$,
which is very small for usual system sizes.
This result can be obtained 
using a matched asymptotics expansion \cite{bazant2004PRE}.
Consequently, we will assume for all $z$, $\rho(z)=0$, or $C(z)=C^+(z)=C^-(z)$
for the perturbation of the charge densities.

ii) Although we do not treat the Debye layers explicitly, we still impose
effective boundary conditions for the electrochemical
potential at the membrane. Thus, we implicitly assume that
the electrochemical potential is continuous across the Debye layers.

We keep the geometry as before, i.e. a flat membrane located at $z=0$ with
given AC voltage $V_0$ at the electrodes located at $z=\pm L/2$.
We again assume that there is no additional
DC field or Nernst potential,
and equal diffusion coefficients\footnote{
Note that in case of unequal diffusion coefficients, one gets
a contribution like $\p_z C$ in Eq.~(\ref{Peqc}).
The subsequent calculations can still be performed in a completely analogous way.
}
for positive and negative ions.
Using the above-discussed approximations, we obtain
\beq
\label{Ceqc}&&i\om C=D\p_z^2 C\,,\\
\label{Peqc}&&\left(i\om\epsilon+D\epsilon\kappa^2\right)\p_z\Phi=I_0\,.
\eeq
Eq.~(\ref{Peqc}) is again easily integrated for $z\in[-L/2,0]$
and together with the BC $\Phi(-L/2)=-V_0/2$
one gets
\beq
\Phi(z)=\frac{I_0(z+L/2)}{i\om\epsilon+D\epsilon\kappa^2}-V_0/2\,.\nonumber
\eeq
In addition we need three more BCs, namely
\beq
\label{cbc1}C(-L/2)&=&0\,,\\  
\label{cbc2}D\left(\p_z C(0)-\bar c\p_z\Phi(0)\right)&=&j^-=0\,,\\
\label{cbc3}D\left(\p_z C(0)+\bar c\p_z\Phi(0)\right)&=&j^+=
\frac{G^+}{e}\left(\frac{k_B T}{c_0}[C]_0+e[\Phi]_0\right)\,,
\eeq
where $[C]_0=C(0^+)-C(0^-)$ and analogously for $[\Phi]_0$.
The second condition is the no-flux condition for the anions.
The third condition states that the bulk current of cations equals
the current through the membrane, and is assumed to follow Ohm's law.
From Eqs.~(\ref{Ceqc}, \ref{cbc1}, \ref{cbc2}), we obtain the following
frequency dependent ion density distribution
\beq
C(z)=\frac{\epsilon\kappa^2 I_0}{2\alpha(i\om\epsilon+D\epsilon\kappa^2)\cosh\left(\alpha L/2\right)}
\sinh\left(\alpha(z+L/2)\right), \nonumber
\eeq
where $\alpha=\sqrt{i\om/D}$ is of Warburg-type, cf.~Eq.~(\ref{beta1Warb}).
Note that here the Warburg impe\-dance arises from breaking the cation/anion symmetry,
due to differences in membrane conductivities rather than due to differences
in their diffusion coefficients as in the previous section.
Also note that although the membrane is non-conductive for the anions,
this is a collective effect in which both types of moving charges participate.

Finally, we use the BC for the cationic current, Eq.~(\ref{cbc3}),
to solve for the voltage $V_0$ and obtain
the impedance via $Z=V_0/(I_0 A)$ as before
\begin{figure}[t]
\centering
\includegraphics[width=0.6\textwidth]{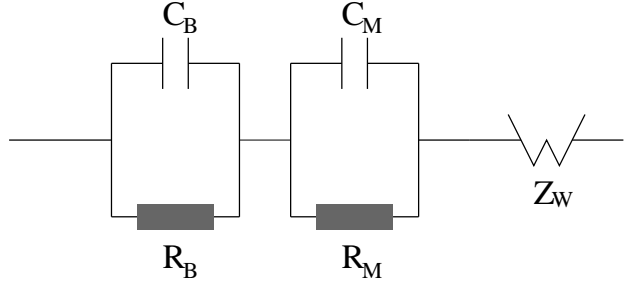}
\caption{\label{fig_ers3}
Effective circuit for the ideally blocking and selectively conductive membrane,
Eq.~(\ref{Zcond}):
Two RC-circuits, one for the bulk and one for the membrane are in series
with a Warburg resistance, caused by the ion selectivity of the membrane.
}
\end{figure}
\beq\label{Zcond}
Z=Z_B   
+\frac{D\kappa^2 /(G^+ A)}{D\kappa^2+i\om}
+\frac{k_B T \kappa^2 /(e c_0 A) }{i\om+D\kappa^2}
\frac{\tanh\left(\alpha L/2\right)}{\alpha}\,.
\eeq
Here we already have identified the bulk circuit, it is present as in the
previous cases.
The second term is the membrane contribution. It can be written as
\beq
Z_M=\frac{1 }{ R_M^{-1}+i\om C_M}\,,
\eeq
with the membrane's resistance, $R_M=1/(G^+ A)$,
and capacitance, $C_m=\frac{G^+ A}{D\kappa^2}$.
The third term
is the Warburg impedance, reading
\beq
Z_W\simeq\frac{2\lambda_D^2}{  D\epsilon A/L   }\frac{\tanh\left(\sqrt{i\omega/D} L/2\right)}{\sqrt{i\omega/D}L}
\eeq
for small $\omega$. Note that it is of the same form as Eq.~(\ref{warburg_2D})
obtained for unequal diffusion coefficients,
except for that in the latter appears an additional
factor containing the diffusion coefficients.

\begin{figure}[t]
\centering
\includegraphics[width=0.6\textwidth]{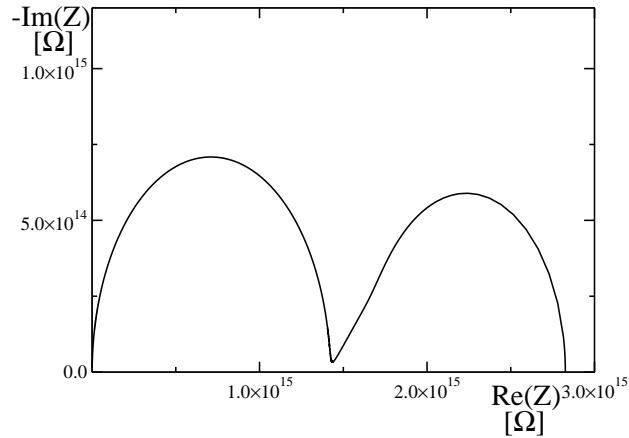}
\caption{\label{fig_cond1}
Nyquist plot for a selectively conducting membrane.
At high frequencies one has
an RC-semi-circle,
which is either dominated by the bulk or by the membrane,
depending on the membrane conductance and the dimensions of the system.
The low frequency behavior is governed by the Warburg impedance.
Parameters as for Fig.~\ref{dispfig} except for: $L=1$mm;
$\kappa=10^{-6}{\rm m}^{-1}$ (pure water); $A=1\mu{\rm m}^2$.
}
\end{figure}

We can conclude that as a result of the ionic membrane selectivity,
a Warburg impedance arises.
Fig.~\ref{fig_ers3} shows the effective circuit.
A Nyquist plot is given in Fig.~\ref{fig_cond1}.
One can identify the typical shape of a Warburg impdeance for low
frequencies: namely, for decreasing frequencies, on leaving the RC-signal of the bulk
$-{\rm Im}[Z(\om)]/{\rm Re}[Z(\om)]$ acquires a slope of $45^o$.
Finally, due to the finite system size ${\rm Im}[Z(\om)]$ vanishes
for $\om\rightarrow 0$.

As already stated above, the calculation in this section \ref{Zconduct}
is oversimplified. By assuming that the electrochemical potential is
continuous across the Debye layers,
there is no explicit contribution from the charging of the Debye layers
to the impedance.
Hence $\lambda_m$, which is important for the charging, does not enter
-- indeed we did not even use the Robin-type condition.
As the membrane is conductive, at least for the cations, charging of the Debye layers
is of minor importance for the overall impedance. With a proper treatment of the
charging of the Debye layers, using a matched asymptotics calculation,
the Robin-type condition will reoccur to match the two solutions and
will reintroduce the length scale $\lambda_m$ into the problem.

\section{Conclusion}

The study and theoretical description of the effects
induced by electric fields on lipid membranes in an electrolyte
is a vast, challenging and far from fully explored problem,
which is of relevance for many applications in biotechnology.

In this review, we have presented a theoretical framework
to understand some of these effects in the simple case of a
planar geometry.
We have seen the importance of capacitive effects,
occurring as a result of charge accumulation in the vicinity of the membrane,
leading to renormalized elastic moduli and to membrane instabilities.
We also have analyzed the flow fields which can be induced
by currents due to small membrane conductivities.
We discussed these effects only for
time-independent (DC) electric fields, i.e.~in the steady-state regime.

Clearly, time-dependent electric fields lead to capacitive charging
of the membrane and to time dependent membrane dynamics.
The capacitive charging can be described in two ways:
the first approach is based on the leaky dielectric
model developed by G. I. Taylor \cite{GItaylor:1969}.
This approach is explained and illustrated in the
contribution of P. M. Vlahovska in the same issue of this book.
One advantage of such an approach is that it captures
the main physical effects associated with capacitive
charging without the complexity of models
which deal explicitly with the ion concentration fields.
For this reason, it is useful to describe
for instance the complex shape changes occurring
in closed lipid vesicles \cite{petia:2009}.

The second approach, which we used in this work,
is based on the electrokinetic Poisson-Nernst-Planck equations.
This more refined level of description includes ion concentration fields,
and therefore it is useful to describe specific effects associated
for instance with the ion transport in ion channels or for effects
occurring in low salt conditions.
It is also needed to describe more precisely the capacitive charging,
which as we have shown here includes several contributions
coming from the bulk, the membrane impedance and the Debye layers themselves.
In this review, we have tried to illustrate the strength
of this approach for quantifying the impedance of a membrane-electrolyte
system. In particular, we have shown how
effective circuits used to interpret experimental data
can be directly derived by this method.
The membrane selectivity with respect
to ion species
is crucial to understand the conduction
properties of membranes with embedded ion channels.
We hope that our work will motivate
further experimental and theoretical investigations in this field.

\section*{Acknowledgements}
We would like to thank particularly Martin Z. Bazant and Petia Vlahovska
for many helpful discussions and access to unpublished work. We would also like
to thank J.~Prost, J.~F.~Joanny, P.~Bassereau, L.~Dinis, G.~Toombes and
S.~Aimon for many inspiring discussions, and the ANR Artif-Neuron for funding.
F.~Z.~thanks the DFG for partial funding via IRTG 1642 Soft Matter Science.

\appendix

\section{Robin-type boundary condition}
\label{app:RobinBC}
In brief, this boundary condition can be motivated for a flat membrane as follows:
since the membrane is assumed to bear no
fixed charges, the normal components of the electric displacement
are continuous at the two interfaces between the membrane and the electrolyte,
\beq\label{contdispl}
\epsilon\p_z\phi(z=\pm d/2)=\epsilon_m\p_z\phi_m(z=\pm d/2)\,,
\eeq
where $\phi_m$ is the electric potential inside the membrane.
Since the electric field $E_m=-\p_z\phi_m$ is constant (to leading order) inside the membrane,
the integral of the inside field can be written in the following way
\beq
E_m d=\int_{-d/2}^{d/2}E_m dz
=-\left[\phi_m(d/2)-\phi_m(-d/2)\right]
=-\left[\phi(d/2)-\phi(-d/2)\right],\nonumber
\eeq
where in the last step we used the continuity of the potential at the membrane surface.
Together with Eq.~(\ref{contdispl}) this yields
\beq
\lambda_m\p_z\phi(z=\pm d/2)=\phi(d/2)-\phi(-d/2)\,.
\eeq
If we take the limit $d\rightarrow 0$ while keeping $\dm=\frac{\epsilon}{\epsilon_m}d$ constant,
one obtains Eq.~(\ref{RobinBC}) in the particular case of $h=0$ and $\mathbf{n}=\hat{z}$. The
same derivation holds for the case of a slightly perturbed membrane surface $h(\rp)$,
where $\rp$ is the in-plane vector.










\end{document}